\documentclass[sigconf]{acmart}

\def\BibTeX{{\rm B\kern-.05em{\sc i\kern-.025em b}\kern-.08emT\kern-.1667em\lower.7ex\hbox{E}\kern-.125emX}}

\usepackage{nicefrac}
\usepackage{siunitx}
\usepackage{array,framed}
\usepackage{booktabs}
\usepackage{
  color,
  soul,
  float,
  epsfig,
  wrapfig,
  graphicx
}
\usepackage{setspace}
\usepackage{latexsym,fancyhdr,url}
\usepackage{enumerate}
\usepackage{algorithm2e}
\usepackage{algpseudocode}
\usepackage{graphics}
\usepackage{xparse} 
\usepackage{xspace}
\usepackage{multirow}
\usepackage{csvsimple}
\usepackage{balance}
\usepackage{graphicx, caption, subcaption}

\usepackage{
  tikz,
  pgfplots,
  pgfplotstable
}
\usepackage{hyperref}

\usetikzlibrary{
  shapes.geometric,
  arrows,
  external,
  pgfplots.groupplots,
  matrix
}

\pgfplotsset{compat=1.9}

\usepackage{mathtools}

\usepackage{xcolor}

\DeclareMathAlphabet{\mathcal}{OMS}{cmsy}{m}{n}

\DeclareGraphicsExtensions{%
    .png,.PNG,%
    .pdf,.PDF,%
    .jpg,.mps,.jpeg,.jbig2,.jb2,.JPG,.JPEG,.JBIG2,.JB2}

\usepackage{xparse}
\newcommand{\bnm}{\begin{newmath}}
\newcommand{\enm}{\end{newmath}}

\newcommand{\bea}{\begin{eqnarray*}}%
\newcommand{\eea}{\end{eqnarray*}}%

\newcommand{\bne}{\begin{newequation}}
\newcommand{\ene}{\end{newequation}}

\newcommand{\bal}{\begin{newalign}}
\newcommand{\eal}{\end{newalign}}

\newenvironment{newalign}{\begin{align}%
\setlength{\abovedisplayskip}{4pt}%
\setlength{\belowdisplayskip}{4pt}%
\setlength{\abovedisplayshortskip}{6pt}%
\setlength{\belowdisplayshortskip}{6pt} }{\end{align}}

\newenvironment{newmath}{\begin{displaymath}%
\setlength{\abovedisplayskip}{4pt}%
\setlength{\belowdisplayskip}{4pt}%
\setlength{\abovedisplayshortskip}{6pt}%
\setlength{\belowdisplayshortskip}{6pt} }{\end{displaymath}}

\newenvironment{newequation}{\begin{equation}%
\setlength{\abovedisplayskip}{4pt}%
\setlength{\belowdisplayskip}{4pt}%
\setlength{\abovedisplayshortskip}{6pt}%
\setlength{\belowdisplayshortskip}{6pt} }{\end{equation}}

\newcounter{ctr}

\newcounter{mytable}
\def\mytable{\begin{centering}\refstepcounter{mytable}}
\def\endmytable{\end{centering}}

\newcounter{myfig}
\def\myfig{\begin{centering}\refstepcounter{myfig}}
\def\endmyfig{\end{centering}}

\newlength{\saveparindent}
\newlength{\saveparskip}
\newcommand{\E}{{\rm I\kern-.3em E}}

\renewcommand{\eqref}[1]{\mbox{Equation~(\ref{#1})}}

\def \part {part}

\renewcommand{\paragraph}[1]{\vspace*{6pt}\noindent\textbf{#1}\;}

\def \blackslug{\hbox{\hskip 1pt \vrule width 4pt height 8pt
    depth 1.5pt \hskip 1pt}}
\def \qed{\quad\blackslug\lower 8.5pt\null\par}

\newcounter{mynote}[section]

\newcommand\ignore[1]{}

\newcounter{rcnote}[section]

\newcounter{mrnote}[section]

\newcounter{fknote}[section]

\newcounter{anote}[section]

\DeclareMathSymbol{\mlq}{\mathord}{operators}{``}
\DeclareMathSymbol{\mrq}{\mathord}{operators}{`'}

\newcommand{\rhf}[2]{R_{f, \gamma}}

\DeclareDocumentCommand{\edist}{o o}{
  \ensuremath{
    \IfNoValueTF{#1}{{d}}{{\sf d}(#1,#2)}
  }
}

\newcommand{\olrk}[1]{\ifx\nursymbol#1\else\!\!\mskip4.5mu plus 0.5mu\left(\mskip0.5mu plus0.5mu #1\mskip1.5mu plus0.5mu \right)\fi}

\NewDocumentCommand{\indseq}{ O{1} O{r} }{{#1}\ldots {#2}}

\setcopyright{none}
\acmConference[ACM NANOCOM '24]{The 11th ACM
International Conference on Nanoscale Computing and
Communication}{October 28--30, 2024}{Milan, Italy}

\begin{document}
\title{A MAC Protocol with Time Reversal for Wireless Networks within Computing Packages}

\author{Ama Bandara}
\affiliation{
\institution{NaNoNetworking Center in Catalunya (N3Cat)\\Universitat Polit\`{e}cnica de Catalunya}
  \city{Barcelona}
  \country{Spain}
  }
\email{ama.peramuna@upc.edu}

\author{Abhijit Das}
\affiliation{
\institution{NaNoNetworking Center in Catalunya (N3Cat)\\Universitat Polit\`{e}cnica de Catalunya}
  \city{Barcelona}
  \country{Spain}
  }
\email{abhijit.das@upc.edu}

\author{F\'atima Rodr\'iguez-Gal\'an}
\affiliation{
\institution{NaNoNetworking Center in Catalunya (N3Cat)\\Universitat Polit\`{e}cnica de Catalunya}
  \city{Barcelona}
  \country{Spain}
  }
\email{fatima.yolanda.rodriguez@upc.edu}



\author{Eduard Alarc\'on}
\affiliation{
 \institution{NaNoNetworking Center in Catalunya (N3Cat)\\Universitat Polit\`{e}cnica de Catalunya}
  \city{Barcelona}
  \country{Spain}
  }
  \email{eduard.alarcon@upc.edu}
 
\author{Sergi Abadal}
\affiliation{
 \institution{NaNoNetworking Center in Catalunya (N3Cat)\\Universitat Polit\`{e}cnica de Catalunya}
  \city{Barcelona}
  \country{Spain}
  }
  \email{abadal@ac.upc.edu}

\date{}

\renewcommand{\shortauthors}{Bandara \textit{et al.}}
\renewcommand{\shorttitle}{A MAC Protocol with TR for Wireless In-Package Communications}

\begin{abstract}

\end{abstract}




\begin{abstract}
Wireless Network-on-Chip (WNoC) is a promising concept which provides a solution to overcome the scalability issues in prevailing networks-in-package for many-core processors. However, the electromagnetic propagation inside the chip package leads to energy reverberation, resulting in Inter-Symbol Interference (ISI) with high delay spreads. Time Reversal (TR) is a technique that benefits the unique time-invariant channel with rich multipath effects to focus the energy to the desired transceiver. TR mitigates both ISI and co-channel interference, hence providing parallel communications in both space and time. Thus, TR is a versatile candidate to improve the aggregate bandwidth of wireless on-chip networks provided that a Medium Access Control (MAC) is used to efficiently share the wireless medium. In this paper, we explore a simple yet resilient TR-based MAC protocol (TR-MAC) design for WNoC. We propose to manage multiple parallel transmissions with simultaneous spatial channels in the same time slot with TR precoding and focused energy detection at the transceiver. Our results show that TR-MAC can be employed in massive computing architectures with improved latency and throughput while matching with the stringent requirements of the physical layer.
\end{abstract}



\maketitle

\section{Introduction}
\label{sec:intro}
The rapid advancement of technology has catalyzed a shift from single-core to multi-core processors and disintegration into multiple \emph{chiplets}, driven by factors such as power constraints, increasing complexity in fabrication, and escalating costs. With the combination of several cores into one single processor, this paradigm shift brings numerous benefits such as increased throughput, enhanced efficiency, and parallel processing capabilities. However, despite these advantages, when the number of cores is increased gradually, communication bottlenecks arise as a challenge that limit the computation performance of the multi-core processors, especially in multi-chiplet systems \cite{Bertozzi2015}. \par 

Currently, Networks-on-Chip (NoCs) and Networks-in-Package (NiPs) \cite{das2024chip,Nip} adapt as a solution for intra- and inter-chip communication, with a set of wired links and integrated routers \cite{Bertozzi2015}. Yet, with the next-generation architectures being scaled towards a large number of cores and across multiple chiplets, NoCs/NiPs encounter significant problems with energy efficiency, especially for long-range and collective communication patterns.\par 

Wireless Network-on-Chip (WNoC) \cite{abadal2024electromagnetic, Shamim2017} is a promising technology that complements wired interconnects with electromagnetic (EM) radiation with integrated antennas and transceivers on cores. In massive multi-core processor architectures, WNoC provides reconfigurability, broadcast, and multicast capabilities with reduced latency. However, to ensure reliability, wireless networks should comply with stringent requirements such as BER$<10^{-12}$, latency of $1-100$ ns, and a throughput of $10-100$ Gbps while meeting low area and power constraints. \par

Wireless communication within a computing package shows distinct challenges. Since EM radiation is confined within the enclosed environment of the chip, it is highly susceptible to reverberation. This reverberation can lead to a significant delay spread due to multipath effects and cause Inter-Symbol Interference (ISI) in high data rates. Moreover, as the energy is dispersed within the package due to reverberation, parallel communications could be hindered by Co-Channel Interference (CCI). Nevertheless, the wireless channel inside the chip package is static and can be pre-characterized. This unique knowledge has paved the way to adapt the technique Time Reversal (TR) \cite{lerosey2004time, alexandropoulos2022time, Han2012}, which creates a spatial matched filter by focusing the energy in both space and time.\par 

TR uses the time-reversed pre-characterized Channel Impulse Response (CIR) to transmit the modulated data to the target receiver. As the on-chip channel is time-invariant, TR achieves a near-perfect concentration at the receiver both in time and space, therefore mitigating both ISI and CCI at the same time, enabling high-speed parallel communications at the same time and frequency \cite{rodríguezgalán2024}.  
Still, in enclosed scenarios like a chip package, the channels between different pairs of wireless nodes will be correlated to some extent. This means that TR will not completely eliminate the CCI; hence, the Number of Parallel Transmissions (NPT) achievable with TR could be limited and will depend on the particular package environment at hand. Yet since the channels can be pre-characterized, the most suitable NPT can be decided up front; and since the wireless channel is time-invariant, the decided NPT will remain static throughout the system's lifetime.\par

In any case, TR does not eliminate the possibility of multiple transmissions interfering with each other. In particular, a collision might occur if two transmitters try to access to the same receiver and a means to avoid or amend such collisions is needed. Traditionally, Medium Access Control (MAC) protocols have been designed to prevent such situations and to make sure that all nodes can reliably access the medium. This is the main focus of this paper.\par

In WNoC, most works consider an orthogonal resource allocation at the link layer \cite{Vijayakumaran2014, Matolak2012}, where a limited bandwidth is statically allocated among the nodes via time, frequency, and code channels -- thus not requiring a MAC protocol. These architectures are not well-suited for traffic bursts and do not scale well with increasing number of cores due to lack of resources and transceiver complexity. On the other hand, recently, there have been few research works exploiting the broadcast nature of WNoC to develop adaptive MAC protocols for the chip environment. In \cite{BRS}, a protocol is proposed based on carrier sensing, the use of collision detection and a negative acknowledgment (NACK) scheme. In the hybrid wired-wireless interconnection network design proposed in \cite{Shamim2017}, as well as in several others works \cite{Mansoor2015, Catania2018}, multiple token passing variants are proposed -- where only the node that holds the token can transmit. 
Another simple yet efficient protocol is explored in \cite{fuzzytoken}, where the merits of token passing and random access variants have been used to build a hybrid protocol to achieve reduced latency at low loads and increased throughput at high loads.\par 

Unfortunately, existing MAC protocols in WNoC are not suitable for TR either because they assume that all nodes listen to all transmissions (which is not the case in TR by definition) or because they statically divide the bandwidth across all nodes, hence missing the opportunity to exploit multiple spatial channels as offered by TR.  
In light of this, in this paper we propose a TR-based MAC protocol (TR-MAC) which is designed by leveraging the merits of TR and to manage the spatial channels while mitigating the CCI due to parallel transmissions. We harness the advantage of using TR in the unique on-chip wireless network to employ multiple spatial channels by improving the latency and efficiency of the overall network along with the aggregate bandwidth. By using orthogonal frequency channels \cite{niknam2016} alongside TR-MAC, effective spectrum sharing can be utilized in massive multi-core architectures, thereby reducing the latency of communication to levels harmonious with computational speed. \par

The rest of the paper is organized as follows. We provide a background on TR communications within the chip package in Section \ref{sec:TR}, while in Section \ref{sec:macdesign} discusses the salient traits of MAC design in WNoC. Section \ref{sec:macprotocol} describes the designed MAC protocol including the basic algorithm with decisions and assumptions. Section \ref{sec:results} analyses the simulation results, while Section \ref{sec:discussion} discusses the possible implementation impairments of TR and its impact on TR-MAC and Sections \ref{sec:conclusion} concludes the paper.

\begin{figure}[t!]
 \centering
\includegraphics[width=0.8\columnwidth]{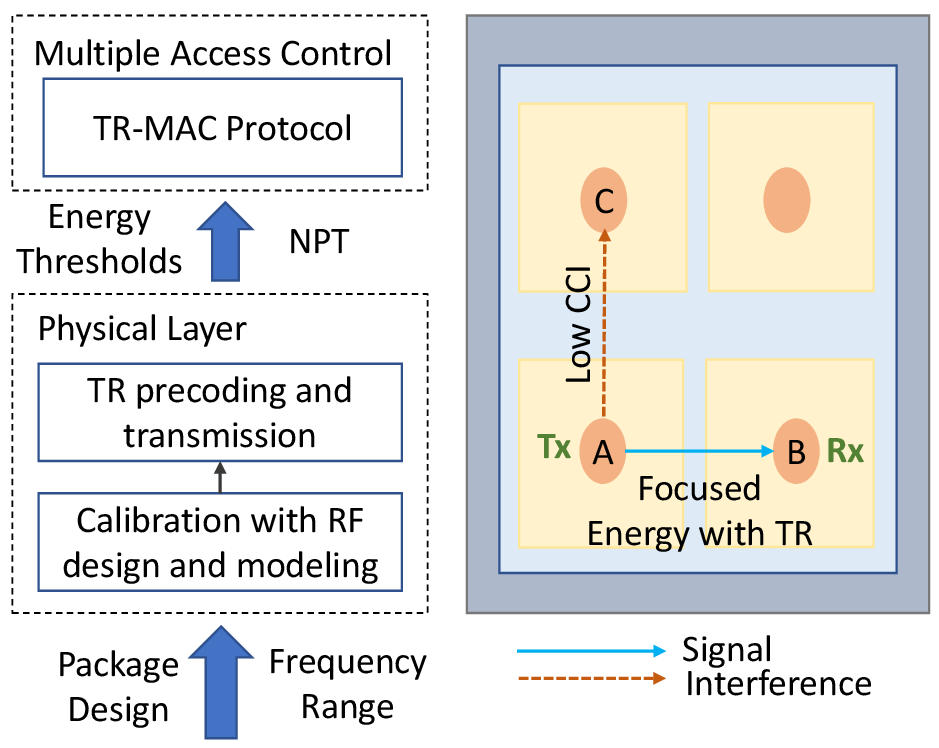}
\vspace{-0.2cm}
\caption{Summary of TR Communication Framework}
\label{fig:framework}
\end{figure}%

\begin{figure*}[t!]
 \centering
\includegraphics[width=1.7\columnwidth,height=5.7cm]{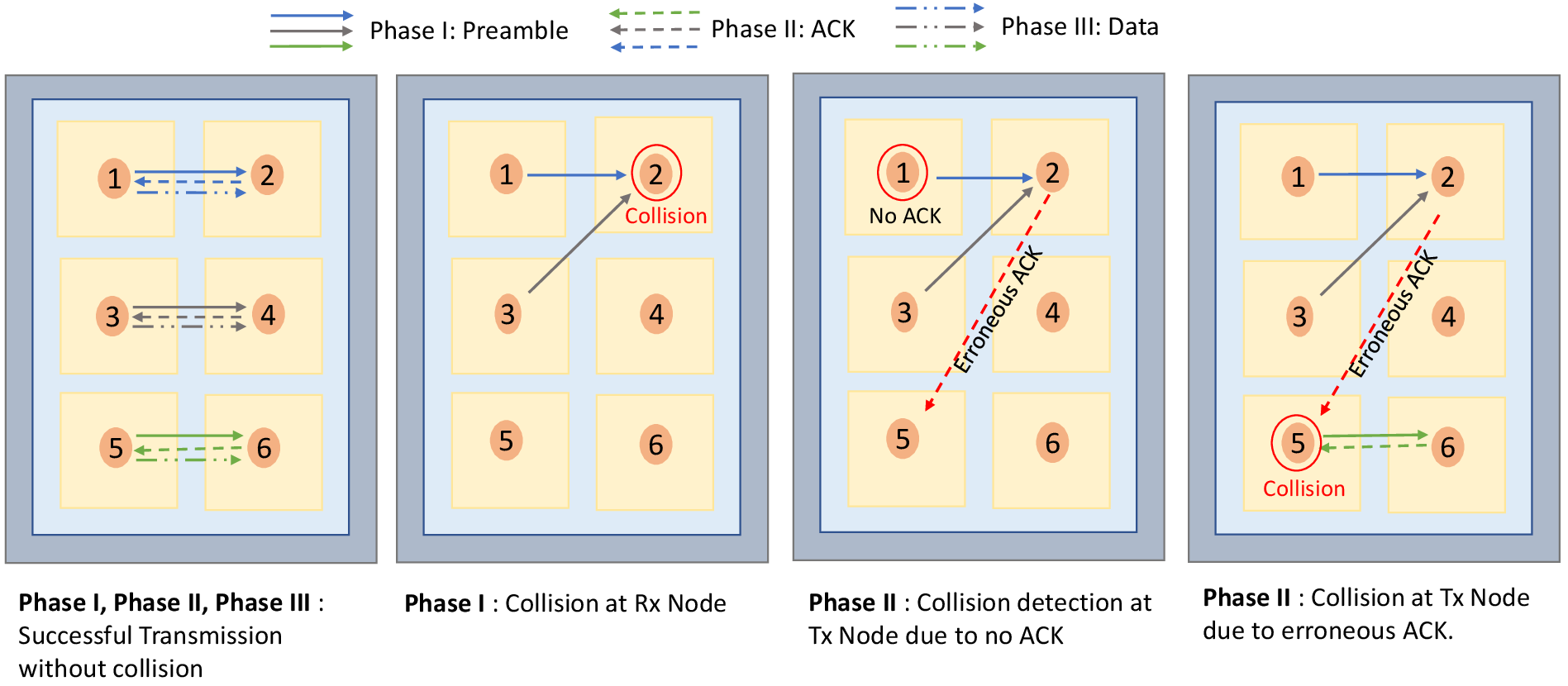}
\vspace{-0.2cm}
\caption{Illustration of TR-MAC Protocol. Leftmost plot represents three successful parallel transmissions without collision. Second plot represents a collision in Phase I due to simultaneous preamble transmissions to node 2. Third plot shows how the collision leads to the original senders not receiving an acknowledgment. Rightmost plot represents the possibility of an erroneous acknowledgment affecting a third transmission.}
\label{fig:collision}
\end{figure*}%

\section{Time Reversal on a Chip Package}
\label{sec:TR}
Wireless communication inside the chip package is subjected to reverberation and thus the EM radiation undergoes high multi-path reflections. The resultant signal at the transceiver has a lengthy delay spread that hinders the performance of high data rate communications causing ISI. Moreover, if multiple concurrent parallel transmissions are to be utilized in WNoC, resource-sharing multiple access methods such as Frequency Division Multiple Access (FDMA), Code Division Multiple Access (CDMA), Orthogonal Frequency Division Multiple Access (OFDMA) or Non-Orthogonal Multiple Access (NOMA) have to be used with the cost of scalability, overheads for data synchronization and complex implementations for an extremely resource-constrained scenario like WNoC.\par

With the benefit of having a rich scattering environment on chip, in our prior work we have explored the TR technique from mmWave-terahertz wireless links on chip for both single and multi-link communications \cite{bandara2023exploration, rodriguez2023}. 
In essence, TR is a technique that uses the pre-characterized CIR to focus the energy perfectly on the intended receiver in both time and space. With unique CIR in between each transmitter and receiver due to multipath effects, the focused energy mitigates both ISI and CCI simultaneously, creating the possibility of collective communication patterns.\par 
 
The basic structure of the TR communication framework is shown in Figure \ref{fig:framework}. There are two phases in TR communication: calibration and transmission. In the calibration phase, CIR is obtained for each transmitter-receiver pair either using simulations or in-situ measurements. The methodology of calibration with full-wave simulations is further explained in \cite{bandara2023exploration}. In the transmission phase, the data is modulated and convoluted with time-reversed CIR and radiated to be detected at the focused receiver. To illustrate a basic TR communication scenario, let us consider a single-link communication between node $A$ and node $B$ with TR, where the neighboring node $C$ is subjected to CCI as shown in Figure \ref{fig:framework}. When node $A$ transmits the TR precoded signal to node $B$ with the use of characterized CIR, node $B$ obtains the perfect energy concentration by maximizing the auto-correlation term of the received signal. Meanwhile, node $C$ follows cross-correlation with the received signal in the spatial domain by minimizing the CCI. As the CIR is time-invariant, the necessary thresholds for reliable data demodulation can be calculated a priori. Finally, there needs to be a mechanism to manage the possibility of having multiple TR transmissions to the same receiver as illustrated in Figure \ref{fig:collision}, since this would lead to catastrophic collisions. This is the purpose of this work.

\section{MAC Design: A Context Analysis}
\label{sec:macdesign}
Wireless communications within package try to improve the latency and reliability of wired NoC/NiP, bounded by area and power constraints, and very much depend on the performance of their MAC protocols. Thus, here we discuss three important verities that contribute to the design requirements of MAC protocol for on-chip scenario. \par

\textbf{Physical Layer.} The benefits offered by the unique landscape of the environment, where the layout and the characteristics of the materials are known beforehand, is leveraged to optimise the communications with high accuracy. The pre-characterized time-invariant channels are analyzed to assess their length, reach consensus on channel allocation based on interference management, and to be aware of impairments such as hidden terminal problems.

\textbf{Architecture.} In high node density, latency plays a crucial role as communication acts as a bottleneck with distinct links. Therefore, the proposed solution has to be simple and minimize time expenditure. 
In addition, the monolithic nature of the system allows us to co-design through the entire architecture, where both ISI and CCI on multiple access are known beforehand and could be effectively managed by designing optimized MAC protocols. 

\textbf{Traffic Traits.} In wireless networks within chip package, traffic can be bursty in time and hotspot in space. Hence, a flexible and easy-to-adapt MAC protocol has to be designed for the highly variable nature of the traffic. Moreover, if multiple access is employed with non-orthogonal resources, it is important to account for high CCI and ensure reliable data transmission between nodes with low error rates.

\section{TR-MAC Protocol}
\label{sec:macprotocol}
The TR-MAC protocol is designed to incorporate the parallel TR transmissions on a single frequency-time channel. The protocol specifically captures the advantages of TR and enhances the performance of WNoC in terms of latency, spectrum sharing and throughput. First, based on physical layer characterizations, the NPT per single frequency channel $n$ has to be predefined, based on the chip environment characteristics and the nature of the channel impulse response.

\subsection{Algorithm}
\begin{figure}[t!]
 \centering
\includegraphics[width=0.98\columnwidth]{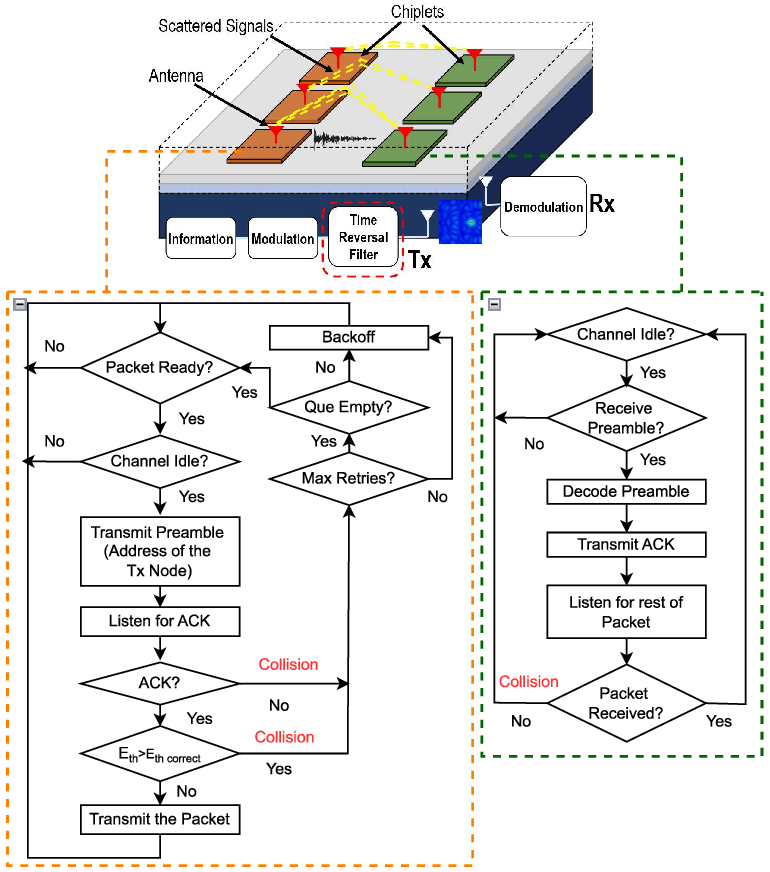}
\vspace{-0.1cm}
\caption{Flowchart of the TR-MAC protocol.} 
\label{fig:algo}
\end{figure}

The algorithm has three phases and is illustrated in Figures \ref{fig:collision} and \ref{fig:algo}. We assume all TR-precoded signal transmissions on physical layer where the signal energy is focused to the intended receivers.

\subsubsection{\textbf{Phase I: Preamble Transmission}}
The node transmits the signal if there is an idle frequency channel assigned for the transmission. If the channels are busy, it waits until a channel is assigned to avoid collision. The transmitter (Tx node) will send a preamble to the interested receiver (Rx node) by using TR. The preamble carries the information of the transmitter address. Hence, by decoding the received preamble, the receiver is able to send an acknowledgment (ACK) of one bit to the Tx node with TR. This way, the ACK will only reach the original transmitter. However, if the Rx node detects the preamble erroneously due to a collision or any other decoding, the acknowledgment will be sent to an incorrect node, which shall ignore it. 

\subsubsection{\textbf{Phase II: Detection of Acknowledgement}}
\label{phase II}
After the preamble transmission in Phase I, the Tx node waits to receive the expected ACK from the Rx node. In this phase, there are two collision scenarios (illustrated in Figure \ref{fig:collision}) that need to be handled.
In the first case, the Tx node does not receive any ACK or detect energy due to a collision that happened in Phase I that led the Rx to send the ACK to someone else. In the second case, the original transmitter detects an ACK with an unexpectedly high amount of energy above a given pre-defined threshold; this is identified as an erroneous ACK. This represents a scenario where two ACKs collide -- one coming from a correct transmission and one erroneous ACK resulting from a collision in Phase I. This occurrence is rare. However, if parallel transmissions occur in same frequency and initially all the transmissions are synchronised in Phase I, there is a low probability of a correlated collision as described here.\par 

Finally, it could be that a node $B$ that is supposed to be Rx of a given transmission from node $A$ starts another transmission at the same time to node $C$. This will be seen as a collision by $A$ because it will not receive any ACK from $B$. On the contrary, if $C$ receives the preamble from $B$ correctly and sends its ACK back, the transmission by $B$ will succeed.

\subsubsection{\textbf{Phase III: Data Transmission}}
If the transmitter detects the correct ACK (the expected threshold of energy concentration of one bit) from the receiver, the preceding data packet is sent to the Rx node.

\subsection{Decisions and Assumptions}
\textbf{Transmitter.} Initially, the available frequency channels are distributed among the nodes as discussed in \cite{multichannel}, following a uniform allocation. Thereby, based on the received energy at the transceiver, the nodes sense the presence of signals, and the collisions are detected. If the transmitter wishes to communicate with more nodes simultaneously by using spatial symmetric channels, the energy detection threshold, $E_{th \, correct}$ is selected with the pre-characterized values of the time-invariant wireless channels and the number of spatial symmetric channels. The received signal at each transmission is compared with $E_{th \, correct}$ to detect the collision at the Tx node. 

\textbf{Receiver.} After the transmission of the preamble on Phase I, which carries the data on the address of the Tx node, the Rx node detects the signal energy with the predefined threshold values. In this scenario, if the received signal is subjected to interference due to multiple attempts to access the same receiver, the demodulated data on the preamble will be an erroneous node address as the Tx node. Therefore, the Rx node will use the time-reversed CIR of another node on the TR filter and the transmission will be erroneous, thus a collision will occur. 

\textbf{Collision Handling and Channel Awareness.} It is assumed that all transmissions are synchronized; i.e., they can only start at the beginning of a global clock cycle. We also assume that 1 cycle is used for preamble transmission and the rest of cycles (4 in this paper) are used for data transmission \cite{bandara2023exploration}. Once a transmission starts, the sender raises a tone through a separate low-bandwidth broadcast channel until the transmission finishes. This technique is a variant of the classical \emph{busy tone} approach and avoids any transmissions to start in the middle of Phase II or Phase III, which could go undetected.

As per the design of the algorithm, when a transmission is initiated, the collisions are handled by Tx node, where the Rx nodes are expected to receive the preamble, re-transmit ACK, and receive data consequently preceding each phase. If a collision happens in Phase II, as described in Section \ref{phase II}, the Tx node goes to exponential back off. For re-transmission, once the node is ready again with a packet, it checks for an idle channel to initiate the retry. Therefore, it will not interfere any existing transmissions on busy channels.\par

\begin{figure*}[h!]
\centering
\begin{subfigure}[t]{0.33\textwidth}
\includegraphics[width=\textwidth]{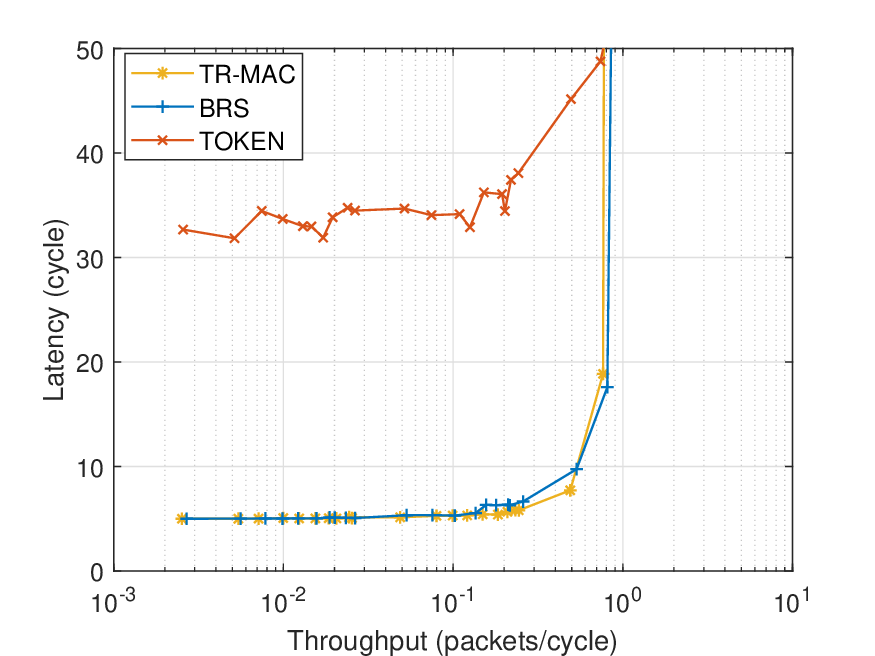}
\caption{}
\label{fig:mulnpt1}
\end{subfigure}
\begin{subfigure}[t]{0.33\textwidth}
\includegraphics[width=\textwidth]{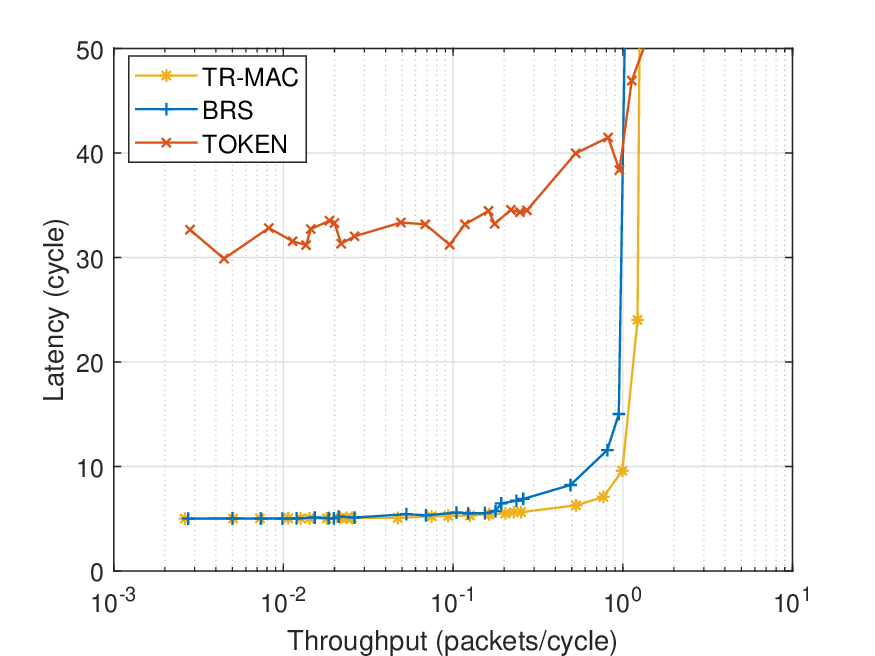}
\caption{}
\label{fig:mulnpt2}
\end{subfigure}
\begin{subfigure}[t]{0.33\textwidth}
\includegraphics[width=\textwidth]{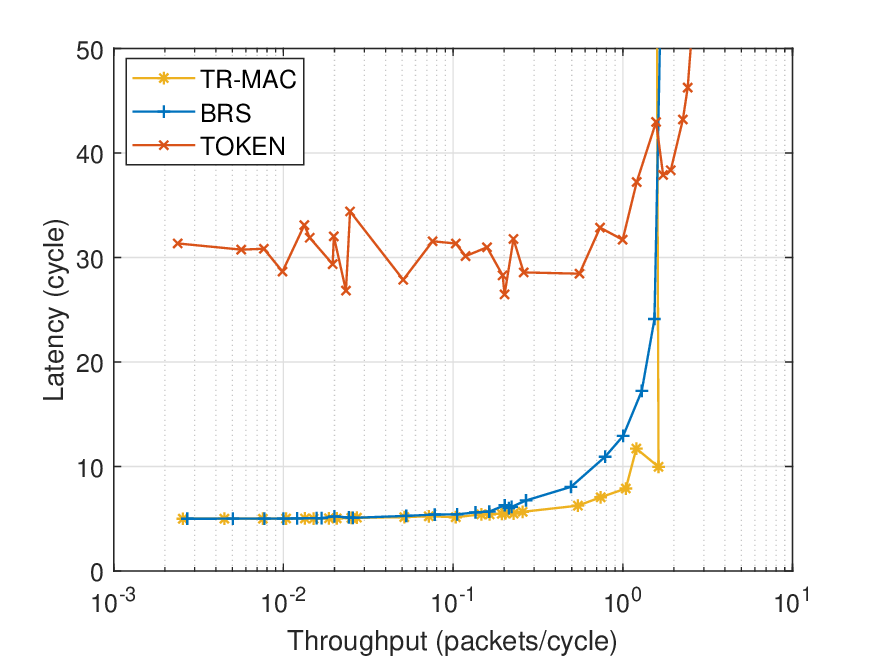}
\caption{}
\label{fig:mulnpt3}
\end{subfigure}
\vspace{0.2cm}
\caption{Latency-throughput curve of TR-MAC and the baseline protocols in an scenario with 64 cores with (a) two channels, (b) three channels, and (c) four channels.} 
\label{fig:npt}
\end{figure*}

\section{Performance Evaluation}
\label{sec:results}

In the following subsections, we evaluate the performance of TR-MAC as a function of the NPT, the number of available channels, and the number of antennas. The simulations were conducted in a modified version of Multi2sim \cite{ubal2012}, which implements a wireless on-chip network. We assume a single link data rate of 30 Gbps \cite{bandara2023exploration} in an scenario with 64 nodes/cores. The synthetic traffic model was characterized as $H=1$ and $\sigma=0.5$, which relates to temporally bursty and moderately hotspot traffic. To compare the performance of the designed protocol, multi-channel versions of random access (BRS) and token passing are considered \cite{multichannel}. 
BRS is a preamble-based MAC that leverages the time-invariant nature of the on-chip wireless medium for collision detection. In token passing, each channel is considered as a token in the shared medium, and each token is allocated to a specific virtual ring created based on the number of cores. In both cases, having $N$ channels imply using $N$ times the amount of spectrum that TR-MAC uses.

\subsection{Number of Parallel Transmissions}
Here, we compare the performance of TR-MAC with different NPT. For a fair comparison, the number of frequency channels in BRS and TOKEN is equal to NPT. 
It is observed in Figure \ref{fig:npt} that TR-MAC achieves a very similar latency than BRS and a saturation throughput comparable to that of TOKEN, but with only a single frequency channel. In low packet injection rates, the number of collisions is low and the latency is almost constant. As the load increases, the network saturates and the latency grows to arbitrarily high values depending on the size of the MAC buffer.

\begin{figure}[h!]
\centering
\includegraphics[width=0.8\columnwidth]{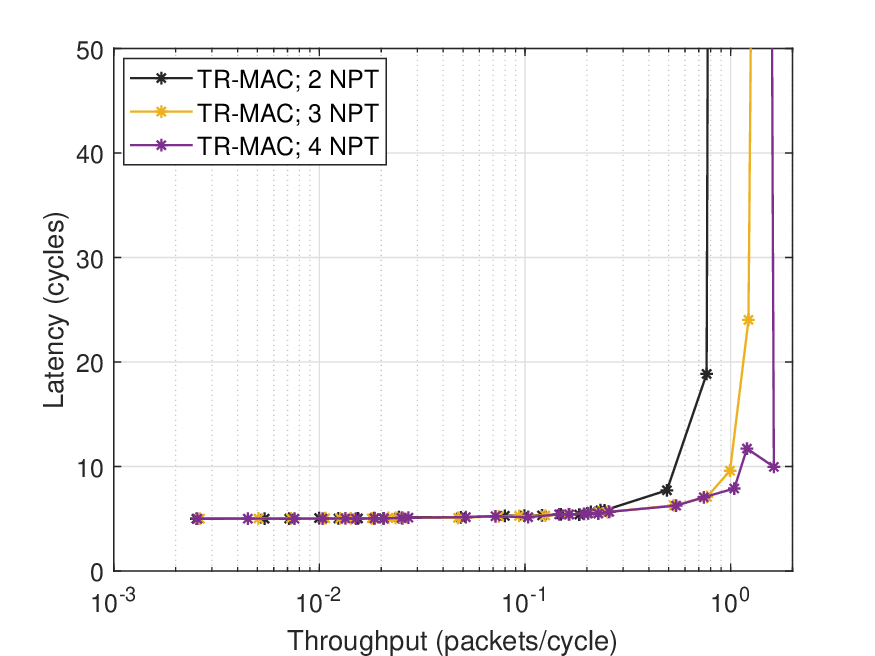}
\caption{Performance of TR-MAC with increasing number of allowed parallel transmissions (NPT) with 64 cores.}
\label{fig:npt1}
\end{figure}

Since with increased number of channels the performance of TR-MAC improves, it is interesting to analyze how NPT affects the overall transmission in terms of throughput and latency. It is shown in Figure \ref{fig:npt1} that with a gradual increment of NPT, the throughput is increased, and the latency is decreased due to the distribution of load with an increased number of transmissions per time slot.\par

\subsection{Number of Channels}
\begin{figure}[h!]
\centering
\includegraphics[width=0.8\columnwidth]{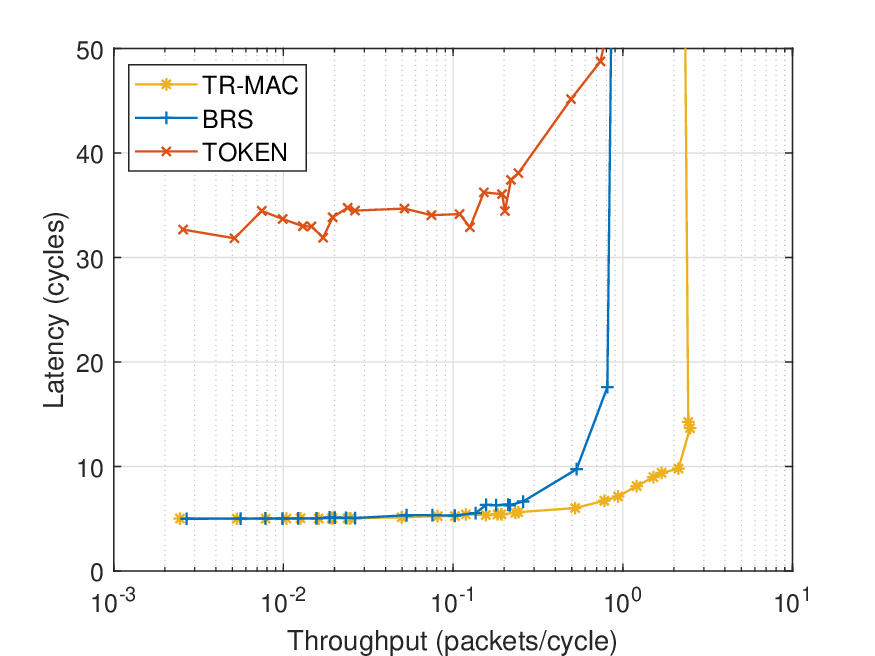}
\vspace{-0.1cm}
\caption{Performance of TR-MAC with NPT=3 on top of two frequency channels in a scenario with 64 cores.} 
\label{fig:numchannels}
\end{figure}

Figure \ref{fig:numchannels} compares the performance of the three evaluated protocols with two frequency channels and, in the case of TR-MAC, three spatial channels (NPT=3). For this, we adapt TR-MAC to choose one of the two frequency channels randomly when transmitting. 

Since TR-MAC accounts for more spatial-frequency channels than the rest, the performance is clearly superior, with a low-load latency similar to BRS but with a much higher throughput -- even higher than token. As latency plays a crucial role in WNoC, we can conclude that TR-MAC provides an efficient and adaptable solution for wireless traffic with superior performance than non-TR counterparts.


\begin{figure*}[t!]
\centering
\begin{subfigure}[t]{0.38\textwidth}
\includegraphics[width=\textwidth]{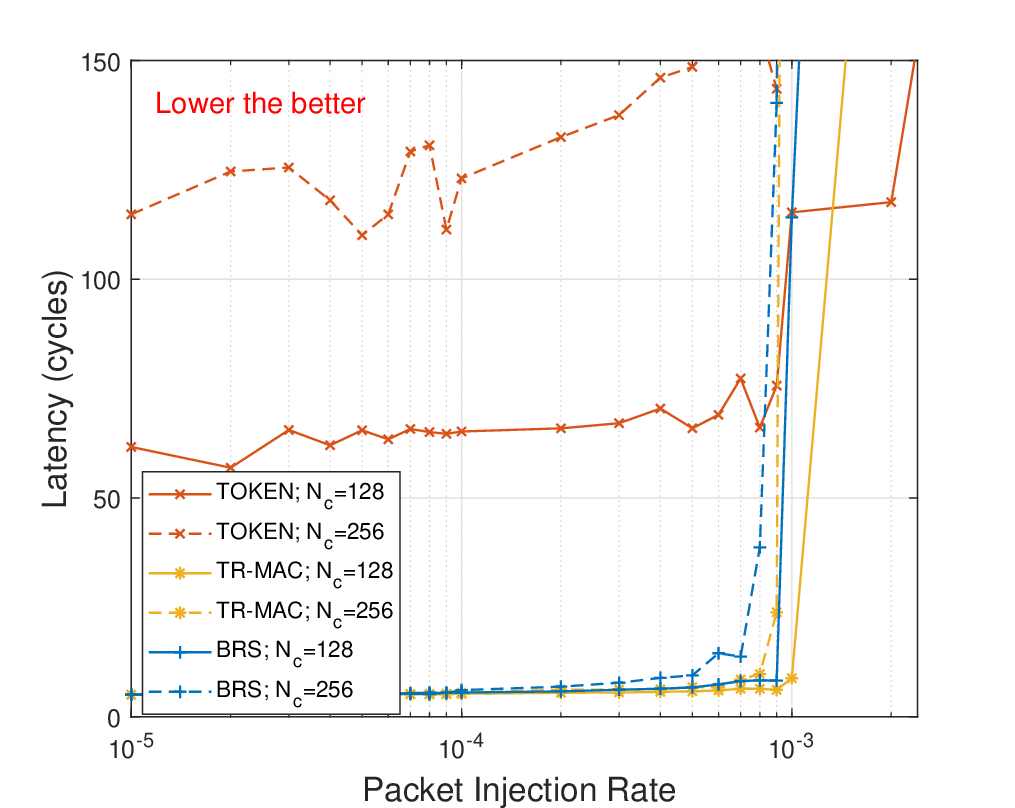}
\caption{}
\label{fig:latency}
\end{subfigure}
\begin{subfigure}[t]{0.38\textwidth}
\includegraphics[width=\textwidth]{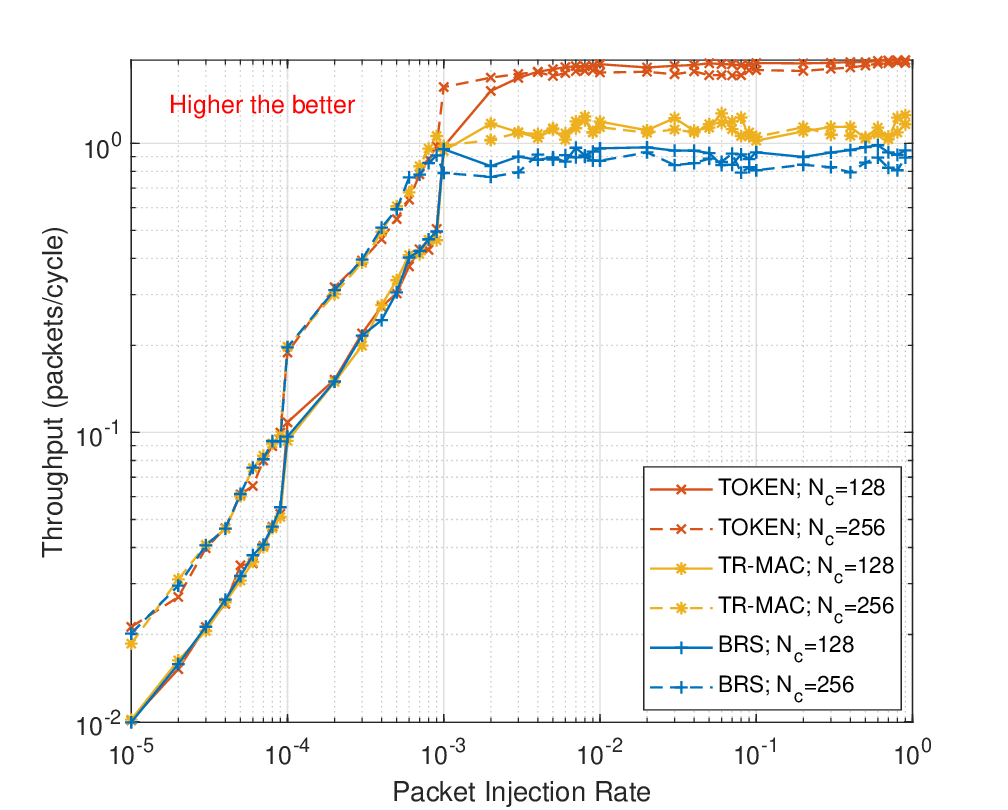}
\caption{}
\label{fig:throughput}
\end{subfigure}
\vspace{0.2cm}
\caption{(a) Latency and (b) throughput of the evaluated MAC protocols for different number of cores with NPT=3 and a single frequency channel as a function of the packet injection rate.}
\label{fig:numcores}
\end{figure*}

\subsection{Number of Cores}
Next, the performance of the protocol is compared based on an increasing number of cores, 128--256, by considering 2 frequency channels for BRS and token passing and NPT=2 for TR-MAC. When the number of cores are increased, if we remain the load of each core fixed, then the traffic is increased proportionally to the number of cores. Therefore, the collision rate increases. Moreover, in token passing, the virtual rings also increase in size and the latency of reaching the interested transmitter also grows. As a result of this, as shown in Figure \ref{fig:latency}, TR-MAC  achieves a lower latency compared with BRS and token passing. However, in terms of throughput (Figure \ref{fig:throughput}), token passing performs slightly better at the expense of a much higher latency compared to both BRS and TR-MAC. This further confirms that TR-MAC stands as a promising solution to achieve high performance (low latency and high throughput) with less frequency resources than other protocols.


\section{Discussion}
\label{sec:discussion}
The TR-MAC protocol is designed as a solution to improve the aggregate bandwidth of on-chip networks by incorporating parallel wireless transmissions in both time and space. Thus, TR-MAC allows to have the performance of multiple frequency channels with a single programmable TR filter, minimising the area and power cost. It is modeled as a preamble-based MAC design by avoiding possible collisions with a TR-filter, to distribute the traffic on each transmission while minimising the resource sharing and improving the latency. TR-MAC leverages the benefit of TR-precoding and decoding schemes of conventional wireless communication and adapts to the unique on-chip wireless network by detecting collisions at the physical layer, while improving the aggregate data rate up to approximately 90 Gbps and beyond with reduced latency and a low BER. Note that, the aggregate bandwidth is the sum of data rates of each parallel transmission in the same frequency channel, but separate spatial channels.


As TR transmissions allow the focus of the energy to the intended receiver, the collisions that would occur with the focused energy could lead to over-the-air computations \cite{OTA} in the wireless medium. Rather than suppressing the interference, if the collisions are leveraged to compute the number of transmitters and use the information to communicate with multiple transmitters with symmetric channels, the network performance would be further improved with carefully designed opportunistic MAC protocols.


\section{Conclusion}
\label{sec:conclusion}

We presented TR-MAC, a scalable versatile protocol for massive wireless architectures that is co-designed with physical layer traits. It is shown that with simultaneous spatial energy concentrations achieved via time reversal, TR-MAC can be used to manage multiple transmissions at the same time-frequency window, which allows us to improve the throughput of the system without sacrificing latency or resources. As TR-MAC is designed to adapt the on-chip networks by leveraging the unique wireless medium, in the future the multi-channel MAC could be further explored with over-the-air computing which leverages the non-orthogonal interference in an opportunistic manner to optimize the network performance.

\bibliographystyle{ACM-Reference-Format}
\bibliography{bib3}

\appendix


\end{document}